\documentclass[twocolumn]{aastex631}
\usepackage{amsmath,amssymb}
\usepackage{graphicx}
\usepackage{booktabs}
\usepackage{longtable}
\usepackage{array}

\newcommand{\astar}{a_{*}}
\newcommand{\rg}{r_{\mathrm{g}}}
\newcommand{\relxill}{\textsc{relxill}}
\newcommand{\xspec}{\textsc{xspec}}
\newcommand{\nustar}{\textit{NuSTAR}}
\newcommand{\xmm}{\textit{XMM-Newton}}
\newcommand{\chisq}{\chi^{2}}
\newcommand{\redchi}{\chi^{2}_{\nu}}
\newcommand{\Nsrc}{57}
\newcommand{\Nelig}{36}

\newcommand{\Nbetterpair}{14}
\newcommand{\Nfits}{222}
\newcommand{\Nacc}{75}

\newcommand{\Npatho}{162}

\newcommand{\Nnestfail}{12}
\newcommand{\Nmsruns}{31}
\newcommand{\Nmsbetter}{20}

\newcommand{\Nnesttests}{36}

\newcommand{\Nspikyprof}{24}
\newcommand{\Nbelowfit}{37}

\newcommand{\SXn}{41}
\newcommand{\SXreq}{27}
\newcommand{\SXreqH}{8}
\newcommand{\SXH}{18}
\newcommand{\SXhiRef}{20}
\newcommand{\SXhiWc}{13}

\newcommand{\CVsrc}{41}

\newcommand{\CVgood}{23}
\newcommand{\CVexpl}{9}
\newcommand{\CVinad}{9}
\newcommand{\CVdiscdet}{20}
\newcommand{\CVdiscnot}{21}
\newcommand{\CVgooddet}{9}

\newcommand{\CVaccepted}{5}

\newcommand{\CVaccLower}{1}
\newcommand{\CVaccUpper}{1}
\newcommand{\CVaccDisc}{2}
\newcommand{\CVaccUncon}{1}

\newcommand{\CVnestarms}{172}
\newcommand{\CVnestfail}{8}

\newcommand{\CVptwod}{10}

\newcommand{\CVheadOld}{18}

\newcommand{\CVxcSrc}{41}
\newcommand{\CVxcN}{82}
\newcommand{\CVxcMin}{0.92}
\newcommand{\CVxcMax}{1.43}
\newcommand{\CVxcMed}{1.20}

\newcommand{\CWcon}{4}

\newcommand{\CWsameIsixty}{3}

\newcommand{\CWunconRthree}{2}

\newcommand{\CWokHd}{1}

\newcommand{\CWsameLp}{0}

\newcommand{\CWokLp}{4}

\newcommand{\CWsameBr}{2}

\newcommand{\CWsameRin}{1}

\newcommand{\CWsameLn}{3}

\newcommand{\CWsameXs}{3}

\newcommand{\CWsameAll}{0}
\newcommand{\CWfaceon}{16}
\newcommand{\CWfaceonNotDet}{8}
\newcommand{\CWhdRail}{4}
\newcommand{\CWaccDetP}{4}
\newcommand{\CWedgeon}{7}
\newcommand{\CWgoodNotDet}{14}

\newcommand{\CWoldLost}{14}
\newcommand{\CWoldLostNotDet}{11}
\newcommand{\CWaccMinIncl}{32}
\newcommand{\CWaccMaxIncl}{56}
\newcommand{\CWlitN}{23}
\newcommand{\CWlitAcc}{2}
\newcommand{\CWlitNotDet}{7}
\newcommand{\CWoldHiInc}{20}
\newcommand{\CWnewHiInc}{10}
\newcommand{\CWoldRailHi}{5}
\newcommand{\CWoldRailLo}{4}
\newcommand{\CWoldAccHi}{10}

\newcommand{\CFdetMedPn}{2.9}
\newcommand{\CFnotMedPn}{2.4}
\newcommand{\CFdetMedNu}{7.1}
\newcommand{\CFnotMedNu}{4.6}
\newcommand{\CFmwuPn}{0.24}
\newcommand{\CFmwuNu}{0.25}
\newcommand{\CFrhoNu}{0.10}
\newcommand{\CFrhoNuP}{0.51}
\newcommand{\CFnotBright}{5}

\newcommand{\CIAllk}{99}
\newcommand{\CIAlln}{120}
\newcommand{\CIAllc}{0.82}
\newcommand{\CIAlle}{0.03}
\newcommand{\CINgcfk}{55}
\newcommand{\CINgcfn}{60}
\newcommand{\CINgcfc}{0.92}
\newcommand{\CINgcfe}{0.04}
\newcommand{\CINgcsk}{44}
\newcommand{\CINgcsn}{60}
\newcommand{\CINgcsc}{0.73}
\newcommand{\CINgcse}{0.06}
\newcommand{\CIAllExtk}{101}
\newcommand{\CIAllExtc}{0.84}
\newcommand{\CIAllExte}{0.03}
\newcommand{\CItruthN}{12}
\newcommand{\CItruthMin}{0.96}
\newcommand{\CItruthMax}{1.11}
\newcommand{\CIredMin}{0.87}
\newcommand{\CIredMax}{1.18}
\newcommand{\CIsLowk}{16}
\newcommand{\CIsLown}{30}
\newcommand{\CIsHik}{28}
\newcommand{\CIsHin}{30}
\newcommand{\CIfalseLL}{11}
\newcommand{\CIfalseLLs}{10}
\newcommand{\CIfalseLLn}{60}
\newcommand{\CIifTrue}{32}
\newcommand{\CIifMin}{3}
\newcommand{\CIifMax}{38}
\newcommand{\CIisTrue}{39}
\newcommand{\CIisMin}{24}
\newcommand{\CIisMax}{43}
\newcommand{\CIcorrHi}{0}
\newcommand{\CIMisk}{9}
\newcommand{\CIMisn}{40}

\newcommand{\CIMisfHi}{0}
\newcommand{\CIMisfiMed}{36}

\newcommand{\CIMisfhalfk}{1}
\newcommand{\CIMisfhalfn}{10}
\newcommand{\CIMisfhalfaMed}{0.92}
\newcommand{\CIMisfhighk}{7}
\newcommand{\CIMisfhighn}{10}

\newcommand{\CIMissk}{1}
\newcommand{\CIMissn}{20}
\newcommand{\CIMissHi}{16}
\newcommand{\CIMissiMed}{84}
\newcommand{\CIMissredMin}{1.07}
\newcommand{\CIMissredMax}{1.25}

\newcommand{\CIMisshalfaMed}{0.87}

\newcommand{\CIMisshighaMed}{0.90}
\newcommand{\CIMissUnres}{16}
\newcommand{\CIMisfHalfAbove}{9}

\newcommand{\CSn}{41}
\newcommand{\CSrhoSx}{0.21}
\newcommand{\CSrhoSxP}{0.19}
\newcommand{\CSrhoFe}{0.22}
\newcommand{\CSrhoFeP}{0.17}

\newcommand{\CBEsoFe}{31}
\newcommand{\CBEsoSum}{44}

\newcommand{\CBNgcsFe}{49}
\newcommand{\CBNgcsSum}{85}

\newcommand{\CBNgcfSoft}{72}
\newcommand{\CBNgcfFe}{13}
\newcommand{\CBNgcfSum}{97}

\newcommand{\CBNgcnSoft}{49}
\newcommand{\CBNgcnFe}{7}
\newcommand{\CBNgcnSum}{73}

\newcommand{\CAgridGap}{1.5}

\newcommand{\CASixtyeightqMin}{4.5}
\newcommand{\CASixtyeightqMax}{6.5}
\newcommand{\CANinetyaMin}{-0.1}
\newcommand{\CANinetyaMax}{0.998}
\newcommand{\CANinetyqMin}{3.0}
\newcommand{\CANinetyqMax}{8.0}

\newcommand{\CResoNew}{1211.4}
\newcommand{\CResoGap}{46}
\newcommand{\CResoGrid}{1211.8}
\newcommand{\CResoGridI}{27}
\newcommand{\CResoFitI}{85}
\newcommand{\CResoI}{3}

\newcommand{\CResoNbGap}{86}

\newcommand{\CResoDetOld}{39.8}

\newcommand{\CRpdsNew}{883.4}
\newcommand{\CRpdsGap}{27}
\newcommand{\CRpdsGrid}{883.5}
\newcommand{\CRpdsGridI}{87}
\newcommand{\CRpdsFitI}{57}
\newcommand{\CRpdsI}{87}
\newcommand{\CRpdsAfe}{10}

\newcommand{\CRpdsZero}{8.2}
\newcommand{\CResoRefl}{10^{-6}}
\newcommand{\CRpdsAhi}{-0.29}

\begin{document}

\title{How reliable are automated X-ray reflection spins of supermassive black holes?
A methodological study of 41 AGN with simultaneous \xmm\ and \nustar\ data}

\author[0000-0003-0095-9302]{K.~Balasubramaniam}
\affiliation{RAD@home Astronomy Collaboratory, Bengaluru, India}
\email{karthikedu77@gmail.com}
\correspondingauthor{K.~Balasubramaniam}
\author{H.-K.~Chang}
\affiliation{Institute of Astronomy, National Tsing Hua University,
  101 Section~2 Kuang-Fu Rd., Hsinchu 30013, Taiwan}
\author{{\L}.~Stawarz}
\affiliation{Astronomical Observatory, Jagiellonian University,
  ul.~Orla~171, 30-244 Krak\'{o}w, Poland}

\begin{abstract}
This is a methodological study: we test how reliable black hole spins from an
automated, uniform reflection analysis are, not how spins are distributed.  We
fit one simultaneous \xmm\ EPIC-pn and \nustar\ observation of each of \CVsrc\
AGN with \texttt{relxillCp}, a distant reflector and a warm-Comptonisation soft
excess, using quality criteria fixed in advance, 20 starting points, spin
profiles scanned in both directions and two-dimensional (spin, inclination)
grids.  We test whether the relativistic reflection is detected, and test every
result against four changes to the model, two sensitivity tests, two alternative
models (a high-density disc and a lamppost corona) and simulations.  The disc
reflection is not detected in \CVdiscnot\ of the \CVsrc\ sources, and \CVaccepted\ sources pass
all checks.  Their 90\% spin regions are \CVaccLower\ lower bound (NGC~4051,
$\astar>0.75$), \CVaccUpper\ upper bound, \CVaccDisc\ regions in separate pieces
and \CVaccUncon\ unconstrained profile; none is a two-sided interval.  The
accepted fits lie at \CWaccMinIncl--\CWaccMaxIncl$^\circ$.  No constrained
region keeps its type under all of the inclination limit, the removal of pn data
below 3~keV, and the two alternative models, so the spin depends on how the soft
excess and the corona are modelled.  In two of the accepted sources, including
NGC~4051, the disc component is detected mainly below 2~keV rather than in the
iron band.  Of the \CVheadOld\ sources accepted by a
previous version of this analysis without a soft-excess component, \CWoldLost\ are
not accepted, \CWoldLostNotDet\ because their disc reflection is not detected;
none of the accepted fits is now edge-on (\CWoldAccHi\ of \CVheadOld\
before).  In simulations with known spin the 90\% regions contain the true
spin in \CIAllc\ of cases when the model is correct; fitting the previous model to
spectra simulated from the new one gives spuriously high spins and, for one of
two templates, edge-on inclinations.  We also document the errors of an automated first pass.
The sample is not complete, and we make no statement about the spin
distribution.
\end{abstract}

\keywords{Supermassive black holes (1663) --- Active galactic nuclei (16) ---
X-ray active galactic nuclei (2035) --- Seyfert galaxies (1447)}

\section{Introduction}
\label{sec:intro}

\begin{figure}
\centering
\includegraphics[width=0.95\columnwidth]{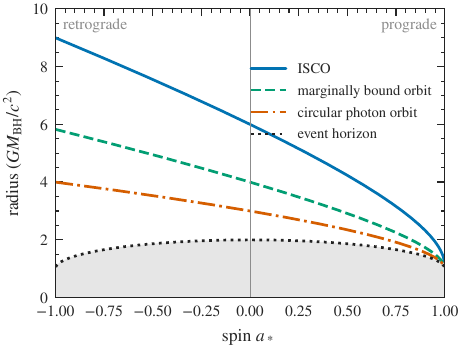}
\caption{Characteristic radii of the Kerr metric in the equatorial plane
\citep{Bardeen1972}.  Reflection spectroscopy measures spin through the ISCO,
assuming that the disc extends down to it.}
\label{fig:kerr}
\end{figure}

The spin of a black hole, $\astar$, sets the radius of the innermost stable
circular orbit (ISCO; Figure~\ref{fig:kerr}; \citealt{Bardeen1972}).  If the
accretion disc extends down to the ISCO, the relativistic distortion of the
X-rays it reflects, in particular the red wing of the Fe~K line and the
Compton hump, depends on $\astar$.  Fitting this reflection spectrum is the
main way in which spins of supermassive black holes have been estimated
\citep{Brenneman2006,Garcia2014,Dauser2014,Reynolds2021}.

Most published AGN spins are high \citep{Reynolds2021}.  Part of the reason is
selection: rapidly spinning, radiatively efficient sources are over-represented
in flux-limited samples \citep{Vasudevan2016}.  Part may be methodological.
Simulations recover the spin reliably only for $\astar\gtrsim0.8$
\citep{Bonson2016}, and blind fits to simulated \xmm+\nustar\ spectra return the
input spin in about half of the cases \citep{Kammoun2018}.  The most recent
compilation lists reflection spins for 51 supermassive black holes, only 22 of
them two-sided; the estimates use different data and models, only about a
quarter rest on quasi-simultaneous \xmm+\nustar\ spectra, and the authors call
for a more uniform sample \citep{SiskReynes2026}.  \citet{Garcia2026} propose
quality criteria for reflection spins, based on whether reflection is detected,
whether the solution is unique, and whether it is robust to reasonable changes
of the model.

This is a methodological paper.  It asks how reliable the spin constraints from
an automated, uniform reflection analysis are, and what such an analysis can and
cannot establish.  It does not attempt to measure the spin distribution of
supermassive black holes; the sample (Section~\ref{sec:sample}) is not designed
for that.  We answer the question in four steps.  First, we fit one simultaneous pn and \nustar\
observation of each source with one reference model, using quality criteria fixed before
the fits were run (Sections~\ref{sec:sample} and \ref{sec:method}).  Second, we
test on simulated spectra whether this procedure returns correct confidence
regions when the model is right, and what it returns when a component is
missing (Section~\ref{sec:injrec}).  Third, we test
whether the relativistic reflection is detected, report the spins, and test each
against changes to the model and against two alternative models of the soft
excess and the corona (Section~\ref{sec:results}).  Fourth, we compare the results with published
reflection spins and with spins obtained by other methods, namely
continuum fitting, jet power and microlensing (Section~\ref{sec:compare}).

The analysis is the corrected version of an earlier automated run of the same
pipeline.  That run produced apparent spin constraints that were the product of
data-handling and fitting errors.  None of its numbers is used here; the errors
are documented in Appendix~\ref{app:audit} because they are easy to repeat.
Uncertainties are 90\% for one parameter ($\Delta\chisq=2.706$) unless stated
otherwise.

\section{Sample and data}
\label{sec:sample}

\subsection{Sample and reduction}
The sample of \Nsrc\ AGN was assembled by hand, in two parts.  The first part
consists of 27 AGN from the reflection-spin compilation of \citet{Reynolds2021}
that have both \xmm\ and \nustar\ observations.  The second consists of 30
further AGN that, at the time the list was made, had archival \xmm\ and
\nustar\ spectra and had been studied with reflection models in the literature;
four of these also appear in \citet{Reynolds2021}, so 31 of the \Nsrc\ sources
are in that compilation (Table~\ref{tab:literature}).  No flux, redshift,
luminosity or classification criterion was applied, and many more AGN have
archival data from both observatories.  The sample is therefore neither complete
nor representative of any AGN population.  It consists mostly of bright, nearby
Seyfert~1s and narrow-line Seyfert~1s, with a few quasars, broad-line radio
galaxies and intermediate or type-2 Seyferts.  For the question asked here this
is adequate: these are the sources to which reflection spin measurements have
been applied, and a test of the method on them is relevant to the published
spins.  The results cannot, however, be extrapolated to AGN in general.

EPIC-pn data were reduced with SAS~22.1.0 ($35''$ source circles,
PATTERN$\le4$, FLAG$=0$, and a flare cut at 0.4~count~s$^{-1}$ in 10--12~keV),
and \nustar\ FPMA/B data with \textsc{heasoft}~6.36 and CALDB 20260608 ($75''$
source circles).  Spectra are grouped to at least 25 counts per bin (pn:
\texttt{specgroup} with oversample 3) and fitted over 0.5--10~keV (pn) and
3--79~keV (FPMA/B).  We use $\chisq$, the abundances of \citet{Wilms2000}, the
cross sections of \citet{Verner1996}, \xspec\ \citep{Arnaud1996} and
\relxill~v2.9.0 \citep{Garcia2014,Dauser2014}.

For pn Small Window observations the background must be taken inside the
read-out window, which is only about $4.4'$ across.  We use a $40''$ circle
(where necessary $25''$) 60--$180''$ from the source, placed where at least 90\%
of the circle is exposed and the count density is lowest; its area is computed
with \texttt{backscale}.  Ten of the 339 Small Window spectra could not be given
a valid background; none of them is used.

Pile-up was checked with \texttt{epatplot} on the source events.  None of the pn
spectra used shows the signature of pile-up, a deficit of single-pixel events
together with an excess of double-pixel events at $>3\sigma$.  Fourteen show a
uniform 3\% excess of single-pixel events, which we attribute to the pattern
calibration.  This rule was refined after the other criteria were fixed: under
the original rule (both ratios within $2\sigma$ of unity) 40 of the 41 eligible
pairs would have been rejected.

\subsection{Choice of observations}
\label{sec:criteria}
For each source we use one pn observation and one \nustar\ observation (a
``pair'').  A pair is eligible if
(i) the two observations start within 2~d of each other,
(ii) each instrument has at least 5000 net counts in its fitted band,
(iii) the background is below half of the source counts,
(iv) the files are complete, and
(v) there is no sign of pile-up.
The 5000-count threshold is a practical choice, not a derived one.  \CVsrc\ of
the \Nsrc\ sources have an eligible pair.  We use the pair chosen by the
automated run (Appendix~\ref{app:audit}) if it is eligible, and otherwise the
best eligible pair in the archive.  The pn
and \nustar\ exposures overlap in time for 38 of these 41 pairs; for Mrk~509,
NGC~7469 and Swift~J2127.4+5654 the gaps are 22, 2 and 10~h.  Fairall~9 and
Ton~S180 have no eligible pair (for Fairall~9 the observations are 127~d apart);
we fit them for comparison only.

\section{Method}
\label{sec:method}

This section describes the model, how it is fitted, how a spin result is
classified, and how its robustness is tested.  The quality criteria were fixed
before any fit of this analysis was run, with the exception of the pile-up rule
(Section~\ref{sec:sample}).  They were informed by the failures of the automated
run (Appendix~\ref{app:audit}), not by the results reported here.

\subsection{Model and fitting}
\label{sec:model}
The reference model is
$\texttt{const}\times\texttt{TBabs}\times\texttt{zxipcf}\times(\texttt{relxillCp}+\texttt{xillverCp}+\texttt{nthComp})$:
Galactic absorption (\texttt{TBabs}, column fixed at the HI4PI value), a
partially covering ionised absorber at the source redshift (\texttt{zxipcf};
column, ionisation and covering fraction free), relativistic and distant
reflection, and a warm-Comptonisation component for the soft excess.

\texttt{relxillCp} provides both the coronal continuum, an \texttt{nthcomp}
spectrum with photon index $\Gamma$ (1.4--3.4) and electron temperature $kT_e$
(5--400~keV), and the relativistically blurred reflection from the disc, whose
strength relative to the continuum is the free reflection fraction (0--10).  Its
spin ($-0.998$ to 0.998), inclination (3--$87^\circ$), ionisation
($0\le\log\xi\le4.7$) and iron abundance (0.5--10) are free; the emissivity
index is fixed at $q=3$, the inner radius at the ISCO, the outer radius at
$400\,\rg$ and the density at $10^{15}$~cm$^{-3}$.

\texttt{xillverCp} describes reflection from distant, neutral material
($\log\xi=0$, density $10^{15}$~cm$^{-3}$) in reflection-only mode.  It is
illuminated by the same continuum: its $\Gamma$, $kT_e$ and iron abundance are
tied to those of \texttt{relxillCp}, and its normalisation is free.  Its
inclination is fixed at $30^\circ$.  In the previous version of this analysis it
was tied to the disc inclination, so the narrow Fe~K$\alpha$ core, which comes
from material far from the black hole, could influence the fitted inclination of
the inner disc, and with it the spin.

The soft excess is modelled by \texttt{nthComp} with $\Gamma$ 1.5--4, electron
temperature 0.05--2~keV and a disc-blackbody seed of 0.01~keV.  The previous
version of this analysis (arXiv v1) had no soft-excess component; adding it to
that model lowered $\chisq$ by more than 11.3 for three extra parameters in
\SXreq\ of \SXn\ sources and changed several spin constraints
(Section~\ref{sec:softexcess}).

\emph{Cross-calibration.}  The multiplicative constant is fixed at 1 for pn and
free, independently, for FPMA and FPMB (allowed range 0.5--1.5).  In the
reference fits of the \CVxcSrc\ eligible sources the \CVxcN\ fitted constants
range from \CVxcMin\ to \CVxcMax, with a median of \CVxcMed; none reaches its
limit, and the FPMA and FPMB values of each source agree closely.  Because the
pn and \nustar\ exposures overlap only partly in time, the constants absorb
source variability between the exposures as well as the flux calibration.  A
difference in spectral slope between the instruments is tested separately
(Section~\ref{sec:variations}).

\emph{Starting points.}  A fit can stop in a local minimum, so each spectrum is
fitted from 20 starting points (five absorber states and inclinations of
$10^\circ$, $30^\circ$, $60^\circ$ and $80^\circ$), and the lowest minimum is
kept.  The spin profile, $\Delta\chisq(\astar)$, is computed on 81 grid points:
41 spaced uniformly over $-0.998\le\astar\le0.998$ and 40 within $\pm0.15$ of
the best fit.  At each point all other parameters are re-fitted, starting from
the solution at the neighbouring point.  The scan is run in both directions;
any point more than $\Delta\chisq=1$ above both neighbours is re-fitted from
each neighbour, and if a lower minimum appears the whole model is re-fitted from
it.

\begin{figure*}
\centering
\includegraphics[width=0.95\textwidth]{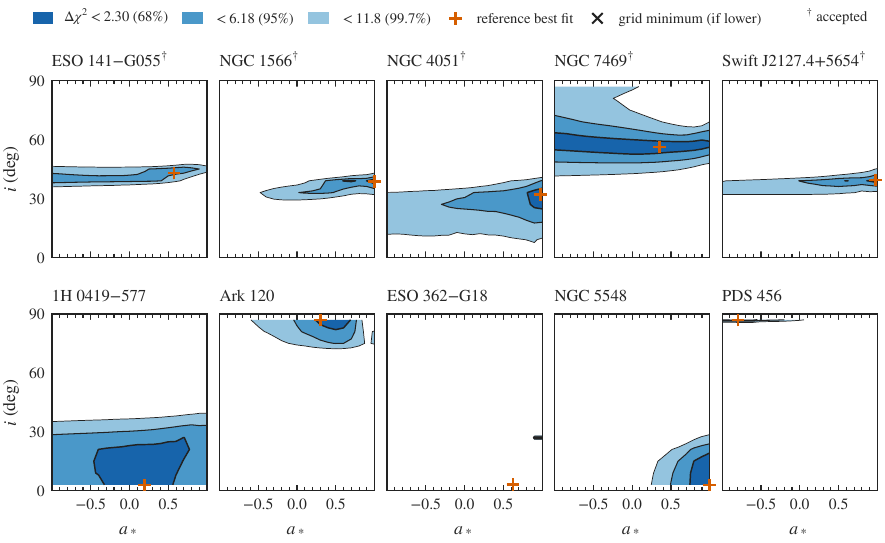}
\caption{Two-dimensional profiles $\Delta\chisq(\astar,i)$ of the reference
model for the \CVptwod\ sources that had an acceptable fit and detected disc
reflection when the grids were run, on a grid of 21 spins and 15 inclinations; all other parameters are
re-fitted at each point, and each row is scanned in both directions.
$\Delta\chisq$ is measured from the lower of the reference fit and the grid
minimum; shaded regions $\Delta\chisq<2.30$, 6.18 and 11.8 (68, 95 and 99.7\%
for two parameters).  Plus: reference best fit (for ESO~362$-$G18 and PDS~456 the fit started from the
grid minimum, Section~\ref{sec:model}); cross: grid minimum where it lies below the
reference fit (none after those refits); $^\dagger$: accepted source.  The grid
of ESO~362$-$G18 was computed from its earlier fit and reaches the new minimum
only near its own minimum; a fit without disc reflection has the same $\chisq$
at every $(\astar,i)$.}
\label{fig:p2d}
\end{figure*}

\emph{Two-dimensional profiles.}  For the \CVptwod\ sources that had an
acceptable fit and detected disc reflection when the grids were run
(Section~\ref{sec:classes}) we also computed
$\chisq$ on a grid of 21 spins and 15 inclinations (Figure~\ref{fig:p2d}) ($3^\circ$ to $87^\circ$ in
steps of $6^\circ$).  Each inclination row is scanned in spin twice, upwards and
downwards, both times starting from the best fit, and the lower value is kept.
In two sources, ESO~362$-$G18 and PDS~456, the grid reached a $\chisq$
lower than the reference fit by more than 1 (by \CResoGap\ and \CRpdsGap), at
$i=\CResoGridI^\circ$ and $\CRpdsGridI^\circ$, while the reference fits lay at
\CResoFitI$^\circ$ and \CRpdsFitI$^\circ$.  Repeating the reference fits from 40
starting points did not reach these minima.  We therefore started a free fit
from the grid minimum itself: the grid row was replayed to recover all
parameters at that point, spin and inclination were then freed, and the spin
profile was recomputed (21 points, inclination free, scanned in both
directions).  Both fits reached the grid minimum ($\chisq=\CResoNew$ and
\CRpdsNew, against \CResoGrid\ and \CRpdsGrid\ on the grid), and these are the
reference fits used in the rest of the paper.  In ESO~362$-$G18 the lower
minimum has no disc reflection (reflection fraction $\sim\CResoRefl$, $i=\CResoI^\circ$,
flat spin profile).  The disc detection at the earlier minimum
($\Delta\chisq_{\rm disc}=\CResoDetOld$) therefore came from a no-disc fit that
had stopped \CResoNbGap\ above the lowest no-disc $\chisq$; disc reflection is not detected in
this source.  In PDS~456 the lower minimum lies at the inclination limit
($i=\CRpdsI^\circ$) with the iron abundance at its upper limit
($A_{\rm Fe}=\CRpdsAfe$) and prefers retrograde spin ($\astar<\CRpdsAhi$ at
90\%; $\Delta\chisq=\CRpdsZero$ at $\astar=0$).  Both $\chisq$ surfaces have
separate minima that 40 random starting points did not find and the
two-dimensional grid did.  The nested changes of Section~\ref{sec:variations} for these
two sources were seeded from the earlier fits and were not repeated; their
\CVnestfail\ fits now lie above the new reference minima.

\subsection{How a result is classified}
\label{sec:classes}
\emph{Spin region.}  The 90\% spin region is the set of grid points with
$\Delta\chisq\le2.706$ above the lowest minimum found.  We report it as one of
five types: a two-sided interval; a lower bound (e.g.\ $\astar>0.72$); an upper
bound (e.g.\ $\astar<0.55$); a region made of separate pieces, which we call
disconnected; or unconstrained (``all'').  An edge within 0.02 of the grid limit
$\pm0.998$ is treated as reaching it.  An upper bound is a real constraint, not
an absence of information.

\emph{Fit quality.}  A fit is \emph{acceptable} if its $\chisq$ probability is
$p\ge0.01$, \emph{exploratory} if $p<0.01$ but $\redchi\le1.2$, and
\emph{inadequate} otherwise.  We do not invoke calibration systematics to admit a
fit.

\emph{Is disc reflection detected?}  A spin can be measured only if the
relativistic reflection is present in the data.  We therefore refit every source
with the disc-reflection fraction fixed at zero, keeping the continuum, the
distant reflection and the soft excess.  This removes four parameters that act
only on the disc reflection (reflection fraction, spin, inclination and disc
ionisation; the distant reflector has its own, fixed inclination).  Disc
reflection is \emph{detected} if $\chisq$ rises by more than 13.28, the 99\%
point of a $\chisq$ distribution with four degrees of freedom.  This threshold
is nominal.  The reflection fraction cannot be negative, which makes the test
conservative, but spin, inclination and disc ionisation are undefined when the
fraction is zero, which can make it anti-conservative; neither case meets the
conditions of the likelihood-ratio theorem \citep{Protassov2002}.  Most detected
sources exceed the threshold by a wide margin (Table~\ref{tab:census}), so the
calibration matters only for the few near it.  The fit without disc reflection is
started from the reference solution only; if it stops above its own minimum, the
detection is overstated, as happened for ESO~362$-$G18 (Section~\ref{sec:model}).

\emph{Warning flags.}  Independently of $p$, a fit receives a flag if
N: it is not at the minimum of its own model (a lower $\chisq$ was found by a
nested model, a profile or the two-dimensional grid);
P: a parameter is at a physically implausible limit (inclination at $3^\circ$ or
$87^\circ$, absorber column at its ceiling, photon index at a limit); or
S: the disc reflection is not detected, or its fraction is at its maximum, so
the disc component carries no spin information.  Parameters at limits that are
physically plausible (iron abundance 0.5, $\log\xi=0$, $kT_e=400$~keV) are
noted but do not disqualify a fit.

\emph{Accepted sources.}  A source is \emph{accepted} if its reference fit is
acceptable and carries no N, P or S flag.  The census below counts only accepted
sources.

\subsection{Robustness tests and alternative models}
\label{sec:variations}
\emph{Changes to the reference model.}  Four changes add freedom to the
reference model, so their best $\chisq$ can only be lower or equal; each is
started from the reference solution:
a broken-power-law emissivity (inner index 3--10 inside a break radius of
2--30~$\rg$, outer index 3);
a free inner radius ($1$--$40\,r_{\rm ISCO}$);
narrow Fe~XXV and Fe~XXVI lines at 6.70 and 6.97~keV that may be in emission or
in absorption; and
a difference in photon index between pn and \nustar\ (a multiplicative
$E^{-\delta}$ term, free for each \nustar\ module within $|\delta|\le0.3$ and
fixed at $\delta=0$ for pn).
Of the \CVnestarms\ such fits, all but the \CVnestfail\ of ESO~362$-$G18 and
PDS~456 (seeded from their earlier reference fits; Section~\ref{sec:model})
reached a $\chisq$ within 1 of the reference minimum or below it, as they should.  Two further tests remove information rather than
add freedom: the inclination is limited to $i\le60^\circ$, and the pn data are
restricted to 3--10~keV.  If a constraint changes under these two tests, it
depends on the inclination prior or on the soft band; this does not by itself
show that it is wrong.

\emph{Alternative reference models.}  The soft excess has two common
interpretations, and the geometry of the corona is not known.  We therefore fit
two alternatives to the whole sample, each from the same 20 starting points:
a high-density disc (\texttt{relxillCp} density free over
$10^{15}$--$10^{19}$~cm$^{-3}$, distant reflector at $10^{15}$~cm$^{-3}$, no
\texttt{nthComp}), in which reflection from a dense disc produces the soft
excess; and a lamppost corona (\texttt{relxilllpCp} with height 1.1--100
horizon radii and a reflection fraction allowed to differ from the
self-consistent lamppost value by a factor 0.33--3, with \texttt{nthComp}).
These are not nested in the reference model.  Where they give a different spin
region, the spin depends on how the soft excess and the corona are modelled.

\subsection{Relation to published quality criteria}
These criteria cover the framework of \citet{Garcia2026}.  Fit quality and the
N, P and S flags correspond to their statistical and model criteria, eligibility
to their simultaneity and instrumental criteria, the detection test to their
detectability criterion, and the changes of Section~\ref{sec:variations} to
their robustness tests.

\section{Does the procedure recover a known spin?}
\label{sec:injrec}

Before applying the procedure to the data, we test it on simulated spectra in
two ways.  First, the spectra are simulated and fitted with the reference model,
which tests the procedure when the model is right.  Second, spectra simulated
from the reference model are fitted with the model of the previous version of
this analysis, which lacked the soft-excess component and tied the inclination
of the distant reflector to that of the disc; this tests what an omitted
component does to the spin and the inclination.

\subsection{Simulations}
The templates are the reference fits of NGC~4051 ($i=\CIifTrue^\circ$) and
NGC~1566 ($i=\CIisTrue^\circ$), two accepted sources with detected disc
reflection.  For each template the spin was set to $\astar=-0.5$, 0, 0.5, 0.8,
0.95 and 0.998, with all other parameters at their best-fit values, and ten
spectra were drawn for each spin with Poisson noise in source and background,
using the real responses, backgrounds and exposures.  The spectra are drawn from
ungrouped copies and given the observed grouping afterwards; drawing them
directly from grouped spectra misplaces counts in energy.  Before any fit,
\CItruthN\ simulated spectra compared with their input model give
$\redchi=\CItruthMin$--$\CItruthMax$; the fits were run only after this check.

Each simulated spectrum was fitted as the real data were: twenty starting points
(five absorber states and four inclinations from 10 to $80^\circ$), taken from
the default values and not from the truth, followed by a profile scanned in both
directions on 41 uniform grid points (step 0.05), with the same re-fitting rules
and the same classification.  The fitted spectra give
$\redchi=\CIredMin$--$\CIredMax$.  For the second test, ten spectra per template
were drawn at $\astar=0.5$ and at 0.95 from the reference model and fitted with
the previous reference model, with the same starting points and profile.

\subsection{Coverage when the model is correct}
\begin{figure*}
\centering
\includegraphics[width=\textwidth]{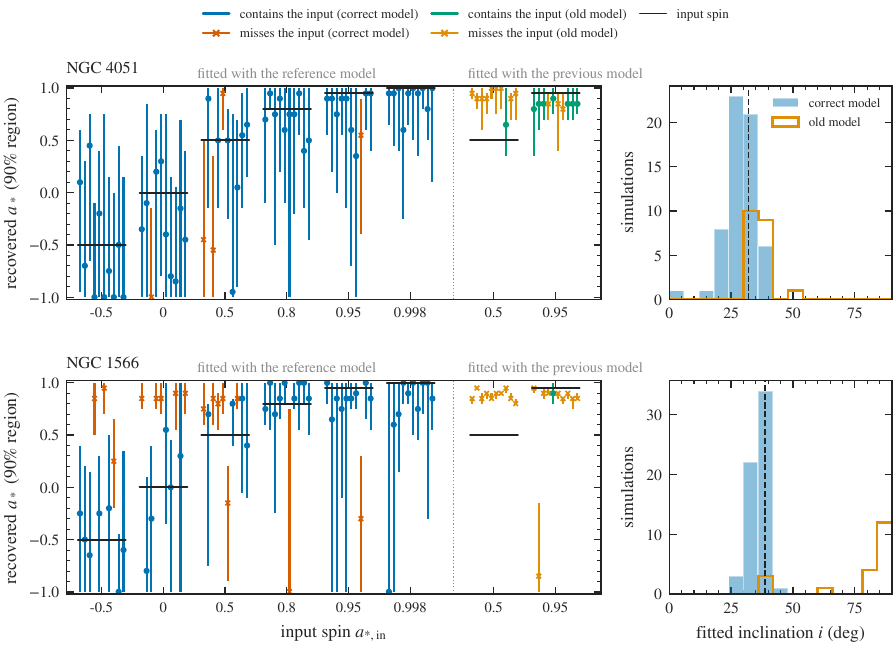}
\caption{Simulations with known spin, for the NGC~4051 (top) and NGC~1566
(bottom) templates.  Left: each vertical bar is the 90\% spin region of one
simulated spectrum (symbol: best fit), grouped by input spin (horizontal black
lines).  Left of the dotted line, spectra fitted with the reference model
(blue: region contains the input; red, crosses: it does not); right of it,
spectra simulated at $\astar=0.5$ and 0.95 and fitted with the previous model
(green: contains the input; orange, crosses: does not).  For disconnected
regions the bar spans the piece containing the best fit.  Right: fitted
inclinations with the reference model (filled) and the previous model
(outline); dashed line: input inclination.}
\label{fig:injrec}
\end{figure*}

Figure~\ref{fig:injrec} shows the recovered regions.  Of the \CIAlln\
simulations, \CIAllk\ have a 90\% region that contains the input spin, a coverage
of $\CIAllc\pm\CIAlle$ (68\% Wilson interval).  The two templates differ: NGC~4051
gives $\CINgcfc\pm\CINgcfe$ (\CINgcfk/\CINgcfn), consistent with 0.90, and
NGC~1566 gives $\CINgcsc\pm\CINgcse$ (\CINgcsk/\CINgcsn).  The shortfall of
NGC~1566 is at low and intermediate spin: \CIsLowk\ of the \CIsLown\ simulations
with $\astar\le0.5$ contain the input, against \CIsHik/\CIsHin\ for
$\astar\ge0.8$.  The grid explains little of it: the edges of a region are the
last grid points inside it, and the input spins $-0.5$, 0.5, 0.8 and 0.95 lie
0.001--0.002 beyond grid points, but widening every edge by one step only raises
the coverage to $\CIAllExtc\pm\CIAllExte$ (\CIAllExtk/\CIAlln).  For a
disconnected region only the piece containing the best fit is counted, which
can only lower the coverage.

Most misses are false indications of high spin: \CIfalseLL\ of the \CIfalseLLn\
simulations with $\astar\le0.5$ give a lower bound that excludes the true spin,
\CIfalseLLs\ of them with the NGC~1566 template.  A lower bound from this
procedure is therefore less secure than its nominal confidence for a source like
NGC~1566.  There is no drift towards edge-on solutions: the fitted inclinations
are \CIifMin--\CIifMax$^\circ$ for NGC~4051 and \CIisMin--\CIisMax$^\circ$ for
NGC~1566, and \CIcorrHi\ lie at $i\ge75^\circ$, although every fit was also
started at $80^\circ$.

\subsection{Fitting the previous model}
\label{sec:misspec}
With the previous model the coverage falls to \CIMisk\ of \CIMisn.  For NGC~1566,
\CIMissHi\ of the \CIMissn\ fits move to $i\ge75^\circ$ (median
\CIMissiMed$^\circ$, true value \CIisTrue$^\circ$), with
$\redchi=\CIMissredMin$--$\CIMissredMax$.  \CIMissUnres\ of them return a spin
region narrower than three grid steps near $\astar\approx0.9$ (median best fit
\CIMisshalfaMed\ for an input of 0.5 and \CIMisshighaMed\ for 0.95), and only
\CIMissk\ contains the input spin.  For NGC~4051 the inclination stays near the
truth (median \CIMisfiMed$^\circ$, \CIMisfHi\ at $i\ge75^\circ$), but at an input
of $\astar=0.5$ only \CIMisfhalfk\ of \CIMisfhalfn\ regions contain it and
\CIMisfHalfAbove\ lie entirely above it (median best fit \CIMisfhalfaMed); at
0.95, \CIMisfhighk\ of \CIMisfhighn\ contain the input.

Omitting the soft excess and tying the distant reflector therefore produces the
two features of the previous version of this analysis, high spins and, for one
of the two templates, edge-on inclinations, in spectra whose true inclination is
moderate and whose true spin is intermediate, and with fits that are not
obviously poor.  This supports the interpretation of
Section~\ref{sec:previous}.

These simulations use two templates, ten spectra per input spin and one
alternative model.  They do not calibrate sources whose spectra the reference
model describes poorly.

\section{Results}
\label{sec:results}

\subsection{Overview}

\begin{figure}
\centering
\includegraphics[width=\columnwidth]{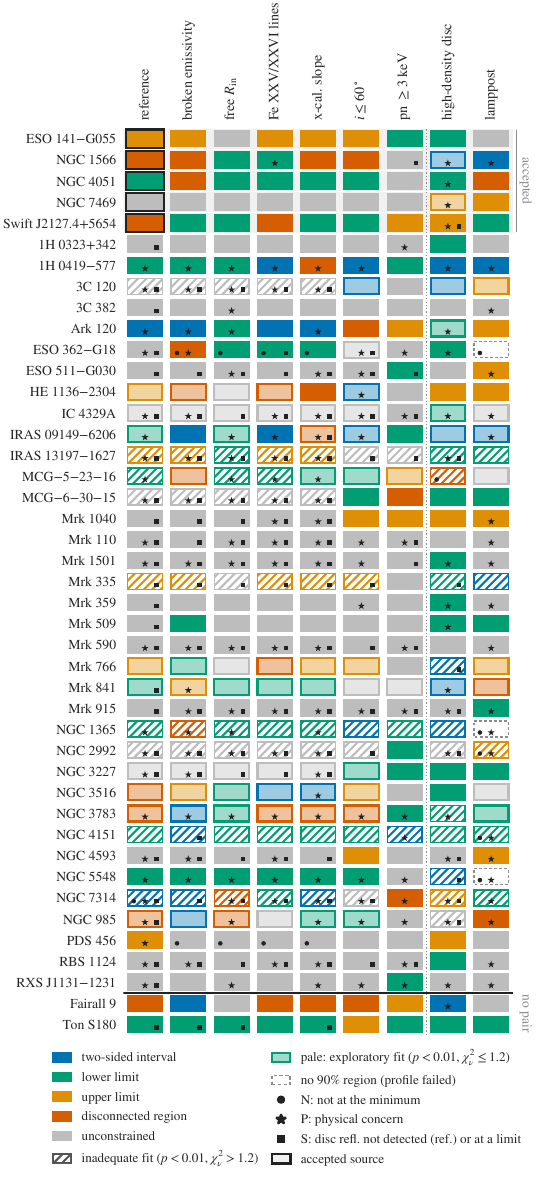}
\caption{The 90\% spin region of each source (rows) under the reference model,
the four nested changes, the two sensitivity tests and, right of the dotted
line, the two alternative models (columns), coded by type.  Hatched: inadequate
fit; pale: exploratory fit; dashed outline: no 90\% region (the profile never
returned to the best fit); circle: N flag; star: P flag; square: S flag (in the
reference column, disc reflection not detected; elsewhere, reflection fraction
at a limit); outlined: accepted sources.  Sources
below the line have no eligible pn--\nustar\ pair.}
\label{fig:matrix}
\end{figure}

Of the \CVsrc\ eligible sources, \CVgood\ reference fits are acceptable,
\CVexpl\ exploratory and \CVinad\ inadequate (Table~\ref{tab:census},
Figure~\ref{fig:matrix}).  Disc reflection is detected in \CVdiscdet\ sources
and not detected in \CVdiscnot; only \CVgooddet\ sources have both an
acceptable fit and detected disc reflection.  Of these, \CWaccDetP\ have
the inclination at a limit (P flag), among them PDS~456 after its fit was
restarted from the two-dimensional grid minimum (Section~\ref{sec:model},
Figure~\ref{fig:p2d}).  ESO~362$-$G18 has left this group: at its lowest minimum
the disc reflection is not detected.

\begin{figure*}
\centering
\includegraphics[width=\textwidth]{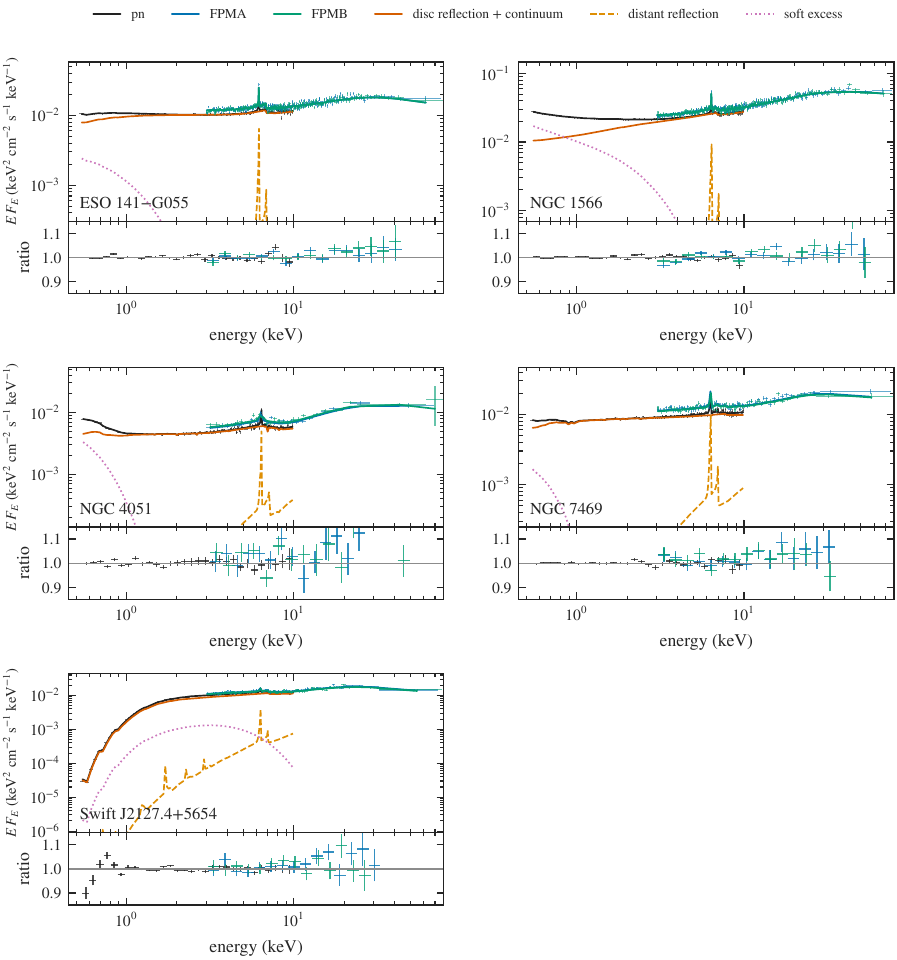}
\caption{Unfolded spectra ($E\,F_E$) of the accepted sources with the reference
model and its additive components (pn), and the data/model ratios.  The spectra
are rebinned for display only (signal-to-noise 15, at most 40 channels per bin);
the fits used the original grouping.}
\label{fig:spectra}
\end{figure*}

The accepted sources, \CVaccepted\ in all, are ESO~141$-$G055, NGC~1566, NGC~4051, NGC~7469
and Swift~J2127.4+5654.  Their spin regions are \CVaccLower\ lower bound,
\CVaccUpper\ upper bound, \CVaccDisc\ disconnected regions and \CVaccUncon\
unconstrained profile (NGC~7469); none is a two-sided interval.  Their spectra
and reference models are shown in Figure~\ref{fig:spectra}.

\begin{figure*}
\centering
\includegraphics[width=\textwidth]{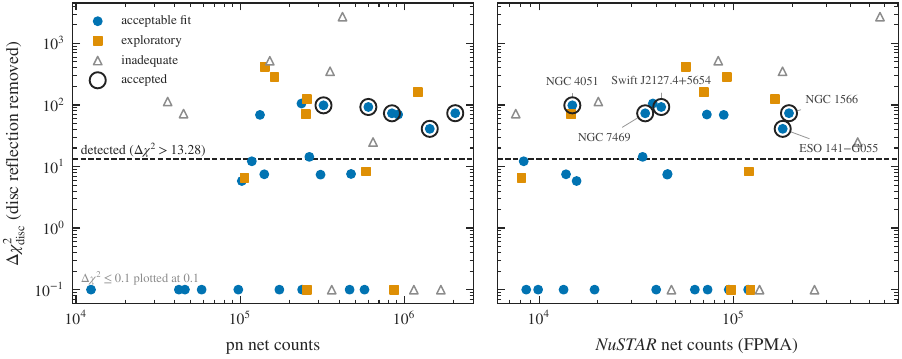}
\caption{Increase of $\chisq$ when the disc reflection is removed from the
reference model, against the net counts of the fitted pn (left) and \nustar\
FPMA (right) spectra, for the \CVsrc\ eligible sources.  Dashed line: detection
threshold ($\Delta\chisq=13.28$, four parameters, 99\%).  Circles: accepted
sources.  Values $\le0.1$ (the disc component does nothing) are plotted at 0.1.}
\label{fig:detect}
\end{figure*}

The most common reason for rejection is therefore not a poor fit but the absence
of measurable disc reflection: \CWgoodNotDet\ sources have an acceptable fit
in which the relativistic component is not detected.  In these sources the
continuum, the distant reflection and the soft excess describe the data, and the
spin is not constrained by the data at all.

\begin{figure*}
\centering
\includegraphics[width=\textwidth]{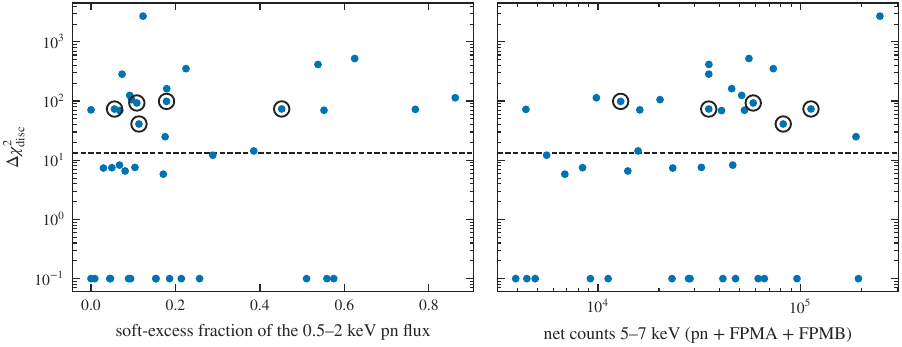}
\caption{$\Delta\chisq_{\rm disc}$ against the fraction of the 0.5--2~keV pn
model flux carried by the soft-excess component (left) and the net counts at
5--7~keV summed over pn, FPMA and FPMB (right), for the \CSn\ eligible sources.
Dashed line: detection threshold; circles: accepted sources; values $\le0.1$ are
plotted at 0.1.  For ESO~362$-$G18 and PDS~456 the soft-excess fractions are
from their original reference fits, not from the fits restarted at the grid
minimum (Section~\ref{sec:model}).}
\label{fig:softexcess}
\end{figure*}

Detection does not simply follow the number of counts (Figure~\ref{fig:detect}).
The sources with and without detected disc reflection have similar median net
counts (pn: $\CFdetMedPn$ and $\CFnotMedPn\times10^5$; FPMA: $\CFdetMedNu$ and
$\CFnotMedNu\times10^4$; Mann--Whitney $p=\CFmwuPn$ and \CFmwuNu), and
$\Delta\chisq_{\rm disc}$ is not correlated with the \nustar\ counts
(Spearman $\rho=\CFrhoNu$, $p=\CFrhoNuP$).  Even among the sources with more than
$5\times10^5$ pn counts, \CFnotBright\ show no detectable disc reflection.  Whether a spin can be
measured therefore depends on the source, not only on the exposure.  Nor does
it follow the strength of the soft excess or the counts in the iron band
(Figure~\ref{fig:softexcess}): $\Delta\chisq_{\rm disc}$ is not correlated with
the soft-excess fraction of the 0.5--2~keV pn flux (Spearman $\rho=\CSrhoSx$,
$p=\CSrhoSxP$) or with the net counts at 5--7~keV ($\rho=\CSrhoFe$,
$p=\CSrhoFeP$).

\begin{figure*}
\centering
\includegraphics[width=\textwidth]{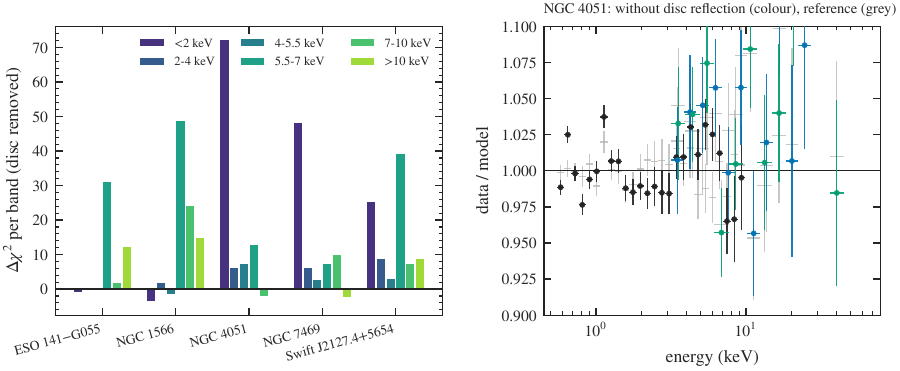}
\caption{Left: increase of $\chisq$ in each energy band when the disc reflection
is removed from the reference model, for the accepted sources (all spectra;
computed as $\sum[(r-1)/\sigma_r]^2$ over the display-binned data/model ratios
$r$, so an approximation to the fitted $\Delta\chisq_{\rm disc}$).  Right:
data/model ratios of NGC~4051 without disc reflection (black: pn; blue: FPMA;
green: FPMB) and with it (grey).}
\label{fig:bands}
\end{figure*}

Where disc reflection is detected, the iron band is not always what detects it.
Figure~\ref{fig:bands} divides the increase of $\chisq$ on removing the disc
component into energy bands, computed from the residuals of the display-binned
spectra; the band sums agree with the fitted $\Delta\chisq_{\rm disc}$ to within
16\%.  For ESO~141$-$G055 and NGC~1566 most of the increase comes from
5.5--7~keV (\CBEsoFe\ of \CBEsoSum\ and \CBNgcsFe\ of \CBNgcsSum) and from the
bands above it, as expected from a broad iron line and a Compton hump.  For
NGC~4051 and NGC~7469 most of it comes from below 2~keV (\CBNgcfSoft\ of
\CBNgcfSum\ and \CBNgcnSoft\ of \CBNgcnSum), and the 5.5--7~keV band contributes
only \CBNgcfFe\ and \CBNgcnFe.  In these two sources the disc component improves
the fit mainly in the soft band, together with the soft excess and the absorber,
rather than through the iron line.  This agrees with the sensitivity test of
Section~\ref{sec:robust}, in which the constraint of NGC~4051 disappears when the
pn data below 3~keV are excluded.

\subsection{Inclination}
\label{sec:inclination}
\begin{figure}
\centering
\includegraphics[width=\columnwidth]{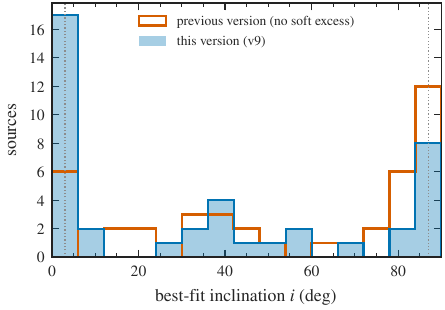}
\caption{Best-fit inclinations of the reference fits of the \CVsrc\ eligible
sources in the previous version of this analysis (no soft-excess component,
distant reflector tied to the disc; outline) and in this version (filled).
Dotted lines: the limits of $3^\circ$ and $87^\circ$.}
\label{fig:incl}
\end{figure}

The accepted sources have inclinations of \CWaccMinIncl--\CWaccMaxIncl$^\circ$,
as expected for Seyfert~1 galaxies; in the previous version of this analysis
\CWoldAccHi\ of its \CVheadOld\ accepted sources lay at $i\ge75^\circ$
(Section~\ref{sec:previous}).  Over all eligible sources the change is smaller
(Figure~\ref{fig:incl}): fits at $i\ge75^\circ$ fall from \CWoldHiInc\ to
\CWnewHiInc\ of \CVsrc, but \CWedgeon\ still lie at the upper limit of
$87^\circ$ (\CWoldRailHi\ before), and fits at the lower limit of $3^\circ$
rise from \CWoldRailLo\ to \CWfaceon.  The inclination remains poorly
determined in most of the sample; what has changed is that the accepted sources
no longer depend on an edge-on solution.

At low inclination the Doppler broadening of the disc line is smallest, so a fit
at the lower limit is one in which the disc contributes little broad emission; \CWfaceonNotDet\ of
these \CWfaceon\ sources also fail the detection test.

\subsection{Robustness and alternative models}
\label{sec:robust}
\begin{figure*}
\centering
\includegraphics[width=\textwidth]{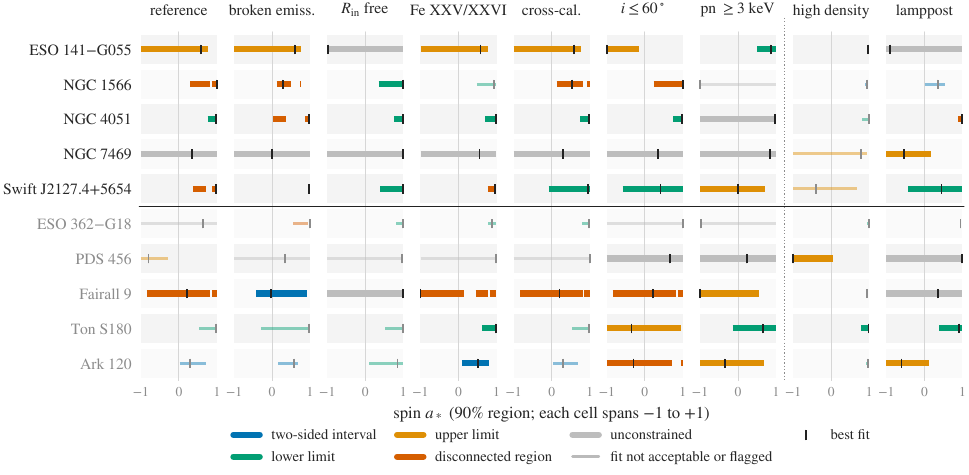}
\caption{Spin regions (90\%) of the accepted sources (above the line) and of the
two sources refitted from their two-dimensional grid minima (ESO~362$-$G18 and
PDS~456; Section~\ref{sec:model}), Ark~120 (inclination at the limit) and Fairall~9 and
Ton~S180 (no eligible pair; grey names) under the reference model, each change and, right of the dotted line,
the two alternative models.  Each cell spans $-1\le\astar\le1$; the colour gives
the type of region and the tick the best fit.  Thin, faint segments: fits that
are not acceptable or carry a flag.}
\label{fig:sens}
\end{figure*}

\begin{figure*}
\centering
\includegraphics[width=\textwidth]{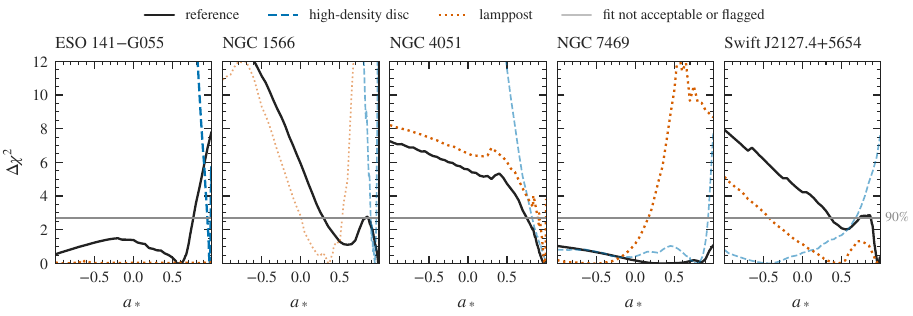}
\caption{Spin profiles $\Delta\chisq(\astar)$ of the accepted sources under the
reference model (solid), the high-density disc (dashed) and the lamppost corona
(dotted), each measured from its own minimum.  Faint lines: fits that are not
acceptable or carry a flag.  Grey line: $\Delta\chisq=2.706$ (90\%).}
\label{fig:prof1d}
\end{figure*}

Tables~\ref{tab:arms} and \ref{tab:arms2} and Figure~\ref{fig:sens} give the
accepted sources under every change of Section~\ref{sec:variations}.  We ask how
many of the \CWcon\ constrained accepted regions keep their type (lower bound,
upper bound, disconnected or two-sided) and whether the new region overlaps the
reference region.

\emph{Changes to the reference model.}  With a broken-power-law emissivity
\CWsameBr\ of the \CWcon\ keep their type, with a free inner radius \CWsameRin,
with Fe~XXV/XXVI lines \CWsameLn, and with a cross-calibration slope \CWsameXs.
In every case the new region overlaps the reference one.

\emph{Sensitivity tests.}  Under $i\le60^\circ$, \CWsameIsixty\ keep their type.
With pn restricted to 3--10~keV, none does and \CWunconRthree\ become
unconstrained: the constraints depend on the soft band, through the continuum and
the soft excess, rather than on the Fe~K line alone.

\emph{Alternative models.}  The high-density disc gives an acceptable fit without
a flag for only \CWokHd\ of the \CVaccepted\ accepted sources, and moves
\CWhdRail\ of them to an inclination limit.  The lamppost model gives acceptable, unflagged
fits for \CWokLp, but \CWsameLp\ of the \CWcon\ constrained regions keep their
type.  No constrained accepted region keeps its type under all four of these
tests (\CWsameAll\ of \CWcon).  The spin found for a given source therefore
depends on how the soft excess and the corona are modelled at least as much as
on the data.  Figure~\ref{fig:prof1d} shows this directly: for NGC~1566 the
reference and lamppost profiles have their minima at different spins, and for
Swift~J2127.4+5654 the high-density disc and the lamppost both prefer lower
spins than the reference model.

\subsection{Individual sources}
\begin{itemize}
\item \emph{NGC~4051} ($i=32^\circ$, $\Delta\chisq_{\rm disc}=98.7$): $\astar>0.75$.
The lower bound is unchanged under $i\le60^\circ$ and with a cross-calibration
slope, and similar with a free inner radius ($>0.77$), narrow lines ($>0.70$), the
lamppost ($[0.89,0.91]\cup>0.92$) and the high-density disc ($>0.82$, but at
$i=3^\circ$).  A broken emissivity adds a low-spin piece
($[0.03,0.36]\cup>0.88$), and the region is unconstrained with pn at 3--10~keV.
With the inner emissivity index $q_1$ free, the two-dimensional profile in
$(\astar,q_1)$ (Figure~\ref{fig:aq}) allows $\astar=\CANinetyaMin$--\CANinetyaMax\
at 90\% confidence for two parameters ($\Delta\chisq\le4.61$, $q_1=\CANinetyqMin$--\CANinetyqMax);
only the 68\% region ($\Delta\chisq\le2.30$) is confined to $\astar\approx0.998$
($q_1=\CASixtyeightqMin$--\CASixtyeightqMax).  This is the most stable constraint in
the sample, but its disc reflection is detected mainly below 2~keV
(Section~\ref{sec:results}), and its high spin depends on the emissivity
assumed.

\begin{figure}
\centering
\includegraphics[width=\columnwidth]{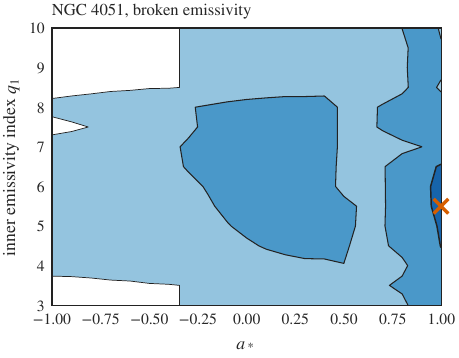}
\caption{Two-dimensional profile $\Delta\chisq(\astar,q_1)$ of NGC~4051 with a
broken-power-law emissivity (outer index 3), on a grid of 21 spins and 15 values
of the inner index $q_1$; all other parameters are re-fitted at each point, and
each row is scanned in both directions.  $\Delta\chisq$ is measured from the free
fit, which lies \CAgridGap\ below the grid minimum; shading: $\Delta\chisq<2.30$,
6.18 and 11.8.  Cross: grid minimum.}
\label{fig:aq}
\end{figure}
\item \emph{Swift~J2127.4+5654} ($i=39^\circ$): $[0.37,0.72]\cup>0.88$.  The
alternatives range from $>0.94$ (broken emissivity) to $<0.67$ (high-density
disc, at $i=3^\circ$) and $<0.72$ (pn at 3--10~keV); the lamppost gives
$>-0.43$.  The spin is not determined.
\item \emph{NGC~1566} ($i=39^\circ$): $[0.29,0.83]\cup>0.86$, nearly unchanged under
$i\le60^\circ$ ($[0.23,0.85]\cup>0.85$).  The lamppost gives $[0.01,0.53]$
(flagged: the corona at the horizon) and the high-density
disc $[0.90,0.97]$ at $87^\circ$ (not acceptable); unconstrained with pn at
3--10~keV.
\item \emph{ESO~141$-$G055} ($i=43^\circ$): $\astar<0.76$, unchanged with a broken
emissivity, narrow lines and a cross-calibration slope, and $<-0.16$ under
$i\le60^\circ$.  With pn at 3--10~keV it becomes a lower bound ($>0.50$), with
the high-density disc $>0.94$ at $80^\circ$, and it is unconstrained in the
lamppost model.
\item \emph{NGC~7469} ($i=56^\circ$): unconstrained in the reference model and in
every nested change; the lamppost gives $<0.17$, and the high-density disc
$<0.93$ at $87^\circ$ (not acceptable).
\end{itemize}

Fairall~9 and Ton~S180, which have no eligible pair, and Ark~120, whose fit is
at the inclination limit, are listed below the line in
Tables~\ref{tab:arms} and \ref{tab:arms2}.  Fairall~9 has an acceptable,
unflagged fit at $33^\circ$ with a region that covers almost the whole range
($[-0.84,0.83]\cup>0.87$).

\subsection{Comparison with the previous version}
\label{sec:previous}
\label{sec:softexcess}
\begin{figure}
\centering
\includegraphics[width=\columnwidth]{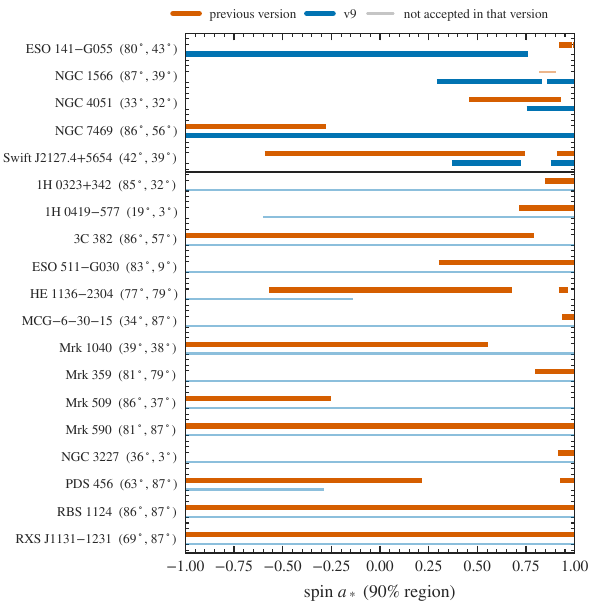}
\caption{Spin regions (90\%) in the previous version of this analysis (upper
bar of each pair) and in this version (lower bar), for every source accepted in
either; thick bars: accepted in that version.  In brackets after each name: best-fit inclinations in the previous version and in this one.
Above the line: accepted in this version.}
\label{fig:v1v9}
\end{figure}

The previous version of this analysis (arXiv v1) used a reference model without
a soft-excess component and with the distant reflector tied to the disc
inclination.  Adding a warm-Comptonisation component to that model lowered
$\chisq$ by more than 11.3 for three extra parameters in \SXreq\ of the \SXn\
sources and in \SXreqH\ of its \SXH\ accepted ones, and reduced the number of
fits at $i\ge75^\circ$ from \SXhiRef\ to \SXhiWc.  With both changes in the
reference model, \CWoldLost\ of the \CVheadOld\ sources accepted in v1 are no
longer accepted, \CWoldLostNotDet\ of them because their disc reflection is not
detected once the soft excess is modelled.  This suggests that many of the
spin constraints of v1, and its edge-on inclinations, came from the disc
component absorbing the soft excess and the narrow Fe~K$\alpha$ core, which
that model had no other way to fit.  Figure~\ref{fig:v1v9} compares the spin
regions of every source accepted in either version.

The simulations of Section~\ref{sec:misspec} test this directly: spectra
simulated from the reference model and fitted with the previous one give
spuriously high spins and, for the NGC~1566 template, edge-on inclinations, with
fits that are not obviously poor.

\section{Comparison with other spin estimates}
\label{sec:compare}

\subsection{Published reflection spins}
Thirty-one of our sources appear in the compilation of \citet{Reynolds2021} and
30 in its 2026 update \citep[][Table~\ref{tab:literature}]{SiskReynes2026}.  Most
published values are lower limits close to maximal spin, obtained with models
built for each source and often with several epochs.  Of the \CWlitN\ that
we fit, only \CWlitAcc\ are accepted in our analysis (NGC~4051 and
Swift~J2127.4+5654), and in \CWlitNotDet\ the disc reflection is not detected
once the soft excess is modelled.

For NGC~4051 our lower bound, $\astar>0.75$, is consistent with the
$\astar>0.99$ listed by \citet{Reynolds2021}; the 2026 update drops that value,
because the spin had been fixed at its maximum in the original analysis
\citep{SiskReynes2026}.  For Swift~J2127.4+5654 our region,
$[0.37,0.72]\cup>0.88$, contains the published $0.72^{+0.14}_{-0.20}$ but is
broader.  For MCG$-$6-30-15, the best-studied case, our reference fit is
inadequate and its disc reflection is not detected, so we cannot compare it with
the published $0.91^{+0.06}_{-0.07}$ or with the $\astar>0.93$ that
\citet{SiskReynes2026} adopt from quasi-simultaneous \textit{XRISM}, \xmm\ and
\nustar\ data (their Table~1 lists $>0.65$).  For Fairall~9 and
Ark~120 the 2026 compilation lists $0.71^{+0.08}_{-0.09}$ and
$0.64^{+0.32}_{-0.06}$.  Our comparison fit of Fairall~9, at $i=33^\circ$, allows
almost any spin ($[-0.84,0.83]\cup>0.87$); that of Ark~120 lies at the
inclination limit.  For Fairall~9 a multi-epoch analysis with an additional soft
Comptonisation component gives $\astar=0.52^{+0.19}_{-0.15}$ at
$i=48^{+6}_{-2}$~deg \citep{Lohfink2012}, while other analyses describe its
X-ray spectrum \citep{Yaqoob2016} and its variability \citep{HagenDone2023}
without relativistic reflection.

\begin{figure}
\centering
\includegraphics[width=\columnwidth]{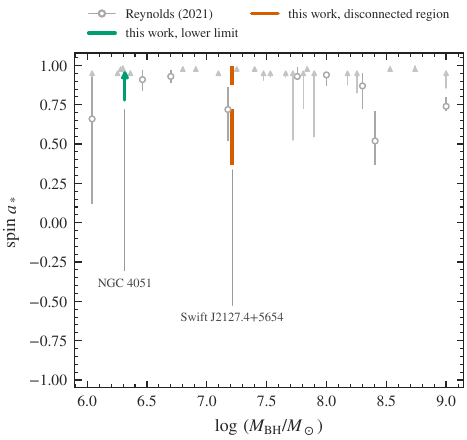}
\caption{Spin against black hole mass.  Grey: \citet{Reynolds2021} (circles,
two-sided 90\% intervals; arrows, lower limits; their masses).  Colour: accepted
sources of this analysis with a mass in that compilation.  No relation is
fitted.}
\label{fig:massspin}
\end{figure}

These comparisons involve different epochs, models and fitting procedures.  They
show that the published constraints are not reproduced by one uniform model on
one epoch; they do not show that the published values are wrong.
Figure~\ref{fig:massspin} places both sets of constraints against the masses
compiled by \citet{Reynolds2021}; we fit no relation.

\subsection{Spins from other methods}
\label{sec:othermethods}
Reflection is not the only route to supermassive black hole spin
\citep[reviewed by][]{Reynolds2021}.  The other methods rest on different
assumptions, so agreement between them is a useful check, although each has
its own systematic uncertainties.

\emph{Continuum fitting} uses the thermal emission of the disc, whose
temperature near the ISCO depends on spin.  In AGN the peak of this emission
falls in the unobservable far ultraviolet, and the method needs the mass,
distance, inclination and accretion rate, so it has been applied to only a few
sources \citep{Done2013,Capellupo2017}.  \emph{Radiative efficiency} arguments
compare luminosity with accretion rate.  Applied to the integrated quasar light
and the local black hole mass density, they imply a mean efficiency that
favours rapid spin on average \citep{Soltan1982,Elvis2002}; applied to
individual quasars they give efficiencies that increase with mass
\citep{DavisLaor2011}, a trend not seen in reflection samples
\citep{Reynolds2021}.  The \emph{outflow (jet-power)} method assumes that the
jet is powered by the spin of the black hole \citep{Blandford1977} and infers
$\astar$ from the jet power, the disc luminosity and the mass \citep{Daly2019}.
\emph{Microlensing} of lensed quasars can magnify the innermost disc and shift
the Fe line, which constrains the ISCO \citep{Chartas2017}.  Horizon-scale
imaging of M87 alone allows $-0.94<\astar<0.94$ \citep{EHT2019V,Reynolds2021};
M87 is not in our sample.

Table~\ref{tab:othermethods} lists the estimates from these methods that we
found for sources in our sample: continuum fits for NGC~3783 and H~1821+643
\citep{Capellupo2017}, a microlensing estimate for RXS~J1131$-$1231
\citep{Chartas2017}, and outflow estimates for 17 sources \citep{Daly2019}.  Of
the \CVsrc\ sources we analyse, 16 have such an estimate; four of the 20 in
Table~\ref{tab:othermethods} have no eligible pair.  We draw four conclusions.

First, only two of these sources are accepted in our analysis.  For NGC~4051
the outflow value, $1.00\pm0.15$, agrees with our $\astar>0.75$.  NGC~7469 is
unconstrained in our reference fit, so its outflow value of 0.94 cannot be
tested.

Second, the conflicts found in the previous version of this analysis have gone.
There, Mrk~509 and NGC~7469 were accepted with edge-on fits ($i=86^\circ$) and
retrograde upper bounds ($\astar<-0.25$ and $<-0.28$), against outflow values of
0.44 and 0.94.  With the soft excess and a separate distant reflector, the disc
reflection of Mrk~509 is not detected and NGC~7469 lies at $i=56^\circ$ with no
spin constraint.  The edge-on, retrograde constraints were therefore a product
of the earlier model, as Section~\ref{sec:previous} suggests.

Third, the continuum and microlensing estimates are too broad to test our
results.  The continuum fit of NGC~3783 gives $0.2^{+0.7}_{-0.9}$ without and
$0.5^{+0.5}_{-0.4}$ with a disc wind \citep{Capellupo2017}; our fit of this
source is exploratory, with its inclination at the lower limit.  For
RXS~J1131$-$1231 the microlensing analysis finds $r_{\rm ISCO}\lesssim8.5\,\rg$
and $i\gtrsim55^\circ$ (99\%) and suggests $\astar\gtrsim0.8$ without a formal
fit \citep{Chartas2017}; in our fit its disc reflection is not detected.

Fourth, the outflow estimates are themselves uncertain for these sources.  The
method assumes that the radio emission traces a spin-powered jet, which is not
established for radio-quiet Seyferts.  It depends on adopted normalisations of
the jet power and the disc luminosity, and values of the spin function above
its maximum are set to $\astar=1$ \citep{Daly2019}, so the six values of 1.00
in Table~\ref{tab:othermethods} are ceilings rather than measurements.  The
agreement for NGC~4051 rests on one source and is not a calibration of either
method.

\section{Discussion}
\label{sec:discussion}

\subsection{What limits the constraints}
Once the data handling and the fits are correct, three things limit the spin
constraints.

The first is whether the disc reflection is in the data at all.  With the soft
excess and the distant reflection modelled separately, the relativistic
component is not detected in \CVdiscnot\ of the \CVsrc\ sources, including
\CWgoodNotDet\ with acceptable fits.  For these sources any spin constraint from
a model without a soft-excess component comes from the disc component fitting
something else: the soft excess, the narrow Fe~K$\alpha$ core, or both.
\citet{SiskReynes2026} likewise note that interpreting the soft excess as
reflection can drive a fit to high-spin lower limits.  A test of whether the
relativistic reflection is required, the detectability criterion of
\citet{Garcia2026}, should precede any spin measurement.

The second is the model of the soft excess and the corona.  For the accepted
sources the high-density disc and the lamppost corona, both physically
motivated, give spin regions that overlap the reference ones but rarely have
the same type (Section~\ref{sec:robust}), and the high-density disc fits most of
them less well.  The spin of a source is therefore not determined by its data
alone but by the data together with a choice of model that the data do not make
for us.

The third is the soft band.  No constrained region keeps its type when only the
3--10~keV pn data are used.  This shows that the constraints depend on the soft
band, not that the soft band is wrong: removing it also removes genuine
reflection information and the leverage on the absorber, the continuum and the
soft excess.

The inclination remains the least reliable parameter.  With the new model the
edge-on fits are halved and none of the accepted sources is edge-on, but \CWedgeon\ of
the \CVsrc\ reference fits still lie at the upper inclination limit and
\CWfaceon\ at the lower one.  A fit at a limit marks a source in which the
inclination is not constrained by the data, usually because the disc reflection
does little; an independent inclination, for example from the narrow-line
region or a jet, would be needed to use such sources.

\subsection{When can a spin from such a pipeline be quoted?}
On the evidence of this analysis, a spin from an automated reflection pipeline
should be quoted only if it comes from an eligible, simultaneous pair; from an
acceptable fit in which the disc reflection is detected; from a fit that is
stable under many starting points and not beaten by a two-dimensional grid;
from a profile scanned in both directions; and together with its behaviour under
alternative models of the soft excess and the corona.  Under these conditions
the pipeline yields \CWcon\ model-dependent constraints among \CVsrc\ sources,
of which only that of NGC~4051 is similar under most changes.  Even when the
model is right, the 90\% regions contain the true spin in \CIAllc\ of simulations
rather than 0.90, and for one template a lower bound often excludes a low true
spin (Section~\ref{sec:injrec}).  In the
terms of \citet{Garcia2026}, none of our constrained regions reaches their
tier~A, which requires stability under plausible model changes.

\subsection{Lessons from the automated run}
The automated first pass of this analysis (Appendix~\ref{app:audit}) produced
apparent constraints, including two-sided intervals for Fairall~9, Ark~120 and
NGC~4051, for reasons unrelated to the black holes: pn backgrounds that were
never subtracted, observation pairs that were months apart or nearly empty, and
fits that had not reached the minimum of their own model.  These errors, not the
number of counts, were the dominant source of misleading results.  Each is
easy to make and hard to see in a single fit, so we recommend that any uniform
analysis check for them explicitly.

The same holds for the final pipeline.  Twenty starting points, and 40 in a
repeat, missed the lowest minimum in two of the \CVptwod\ sources with
two-dimensional grids; only the grid found it, and in one of the two
(ESO~362$-$G18) the lower minimum removes the disc reflection altogether.  A
multistart fit that ends at a high inclination or with a detected disc should
therefore be checked against a grid in $(\astar,i)$ before it is used.

\section{Conclusions}
\begin{enumerate}
\item One uniform reflection analysis of \CVsrc\ AGN with simultaneous pn and
\nustar\ data, with a soft-excess component and a separate distant reflector in
the reference model, accepts \CVaccepted\ sources: ESO~141$-$G055, NGC~1566,
NGC~4051, NGC~7469 and Swift~J2127.4+5654.  Their 90\% spin regions are
\CVaccLower\ lower bound (NGC~4051, $\astar>0.75$), \CVaccUpper\ upper bound,
\CVaccDisc\ disconnected regions and \CVaccUncon\ unconstrained profile; none is
a two-sided interval.  In ESO~362$-$G18 and PDS~456 the two-dimensional grid
found minima that 40 random starting points had missed; at the lower minimum
ESO~362$-$G18 has no disc reflection, and PDS~456 is at the inclination and
iron-abundance limits.
\item The relativistic disc reflection is not detected in \CVdiscnot\ of the
\CVsrc\ sources.  This, not poor fits, is the main reason sources are rejected.
\item The accepted sources lie at $i=$\CWaccMinIncl--\CWaccMaxIncl$^\circ$.
None is edge-on, whereas \CWoldAccHi\ of the \CVheadOld\ accepted in the
previous version were.  Over all sources the fits at $i\ge75^\circ$ are halved
(\CWoldHiInc\ to \CWnewHiInc), but fits at an inclination limit remain
common, now mostly at $3^\circ$.
\item The spin regions depend on the model.  No constrained region keeps its type
under all of: $i\le60^\circ$, pn at 3--10~keV, a high-density disc and a lamppost
corona.  The most stable constraint is NGC~4051, $\astar\gtrsim0.7$ under most
changes, but its disc reflection is detected mainly below 2~keV and, with the
inner emissivity index free, its spin is not constrained at 90\% confidence.
\item Of the \CVheadOld\ sources accepted in the previous version, \CWoldLost\ are
no longer accepted, \CWoldLostNotDet\ because their disc reflection is not
detected once the soft excess is modelled.
\item In simulations with known spin and the correct model, the 90\% regions
contain the input spin in $\CIAllc\pm\CIAlle$ of cases (\CINgcfc\ and \CINgcsc\
for the two templates), with no drift towards edge-on solutions.  Fitting the
previous model to spectra simulated from the reference model lowers this to
\CIMisk\ of \CIMisn, with spuriously high spins and, for one template, edge-on
inclinations.
\item Published reflection spins of the same sources are usually higher and
narrower; where our fit is accepted the comparison is consistent (NGC~4051,
Swift~J2127.4+5654).  The outflow estimate of NGC~4051 agrees with ours; the
conflicts with outflow estimates found for edge-on fits in the previous version
have gone.
\item The automated first pass failed through unsubtracted backgrounds,
mismatched observations and unconverged fits; these, not photon statistics,
dominated its errors.
\item The sample was assembled by hand and is not complete; no statement about
the spin distribution or a spin--jet relation is supported.
\end{enumerate}

\begin{acknowledgements}
K.B.\ acknowledges support from the PETC Overseas Scholarship, Government
of Karnataka, India (award DSTW/ED-1/CR-4/2016-17; 2016--2021).  K.B.'s
postdoctoral fellowship at National Tsing Hua University
(Aug.\ 2022--Jul.\ 2023), and H.-K.C.'s contribution, were supported by
MOST grant 111-2112-M-007-019 (NTHU project 111B0034I9; MOST is now the
National Science and Technology Council, NSTC).  This
work uses observations obtained with \xmm, an ESA science mission with
instruments and contributions directly funded by ESA Member States and NASA,
and data from the \nustar\ mission, a project led by the California Institute
of Technology, managed by the Jet Propulsion Laboratory, and funded by NASA.
Fits were run on the computing cluster of the Astronomical Observatory of the
Jagiellonian University.
\end{acknowledgements}

\facilities{XMM-Newton (EPIC-pn), NuSTAR}
\software{XSPEC \citep{Arnaud1996}, relxill \citep{Garcia2014,Dauser2014},
HEASoft 6.36, SAS 22.1.0, Matplotlib, Seaborn}

\section*{Data availability}
All data are public in the HEASARC and XSA archives.  The audit files, the
analysis manifest, fit results, profiles, scripts and figure data will be
released with the paper.

\appendix
\section{The automated first pass}
\label{app:audit}

The automated pipeline paired each source's pn spectrum with the \nustar\
observation closest in time and fitted
$\texttt{const}\times\texttt{TBabs}\times\textsc{abs}\times(\texttt{relxillCp}+\texttt{xillverCp})$
once, from one starting point, under five absorber and reflector-density
variants; spin was then profiled.  None of its numbers is used in the body of
the paper.  Its failures fall into five groups.

\paragraph{Backgrounds.}  The pn background circle was placed $5'$ from the
source.  In Small Window mode the read-out window is about $4.4'$ across, so the
circle fell outside it.  All 339 Small Window pn spectra on disk, 38 of them in
the selected epochs, therefore have background spectra with zero counts and zero
area, and \xspec\ fitted them without background subtraction.  For Fairall~9 the
missing background is 0.7\% of the 0.5--10~keV counts, 3.6\% at 5--7~keV and
11\% at 8--10~keV.

\paragraph{Choice of observations.}  Only \Nelig\ of the \Nsrc\ selected pairs
meet the eligibility criteria of Section~\ref{sec:criteria}.  Eight pairs are
more than a month apart, five \nustar\ observations are shorter than 500~s, and
five pn spectra contain fewer than 100 counts because the source region did not
contain the source (in three cases also because of severe background flaring).
For \Nbetterpair\ sources a substantially better simultaneous pair exists in the
archive.

\begin{figure*}
\centering
\includegraphics[width=0.9\textwidth]{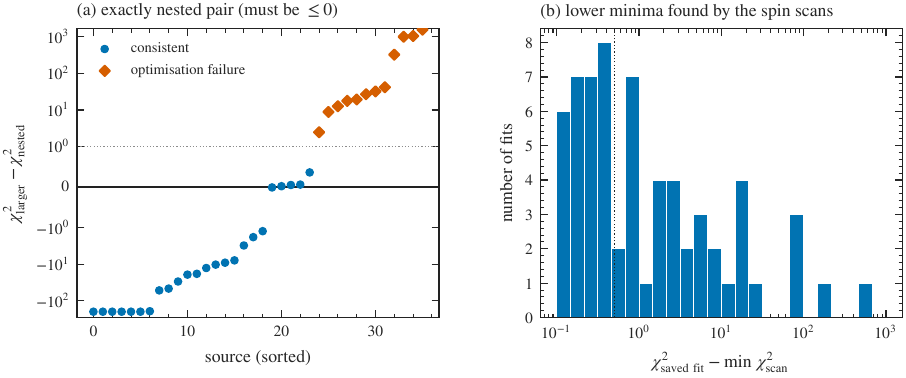}
\caption{Automated first pass: optimisation diagnostics.  Left: $\chisq$ of the
larger of the exactly nested pair minus that of the smaller (positive values are
failures).  Right: improvement of the lowest profile point over the saved fit.}
\label{fig:auditopt}
\end{figure*}

\paragraph{Optimisation.}  In one pair of variants the larger model contains the
smaller exactly (a free disc density tied to the distant reflector contains both
densities at $10^{15}$~cm$^{-3}$), so its $\chisq$ cannot be higher.  Yet
\Nnestfail\ of \Nnesttests\ fits of the larger model ended more than
$\Delta\chisq=1$ above the smaller.  Ten-start refits of the reference model
found a lower minimum than the single-start fit for \Nmsbetter\ of \Nmsruns\
sources, by up to $\Delta\chisq=3459$.  In \Nbelowfit\ profiles a scan point lay
below the saved fit, and \Nspikyprof\ profiles contained isolated failed steps
(Figure~\ref{fig:auditopt}).

\begin{figure*}
\centering
\includegraphics[width=0.46\textwidth]{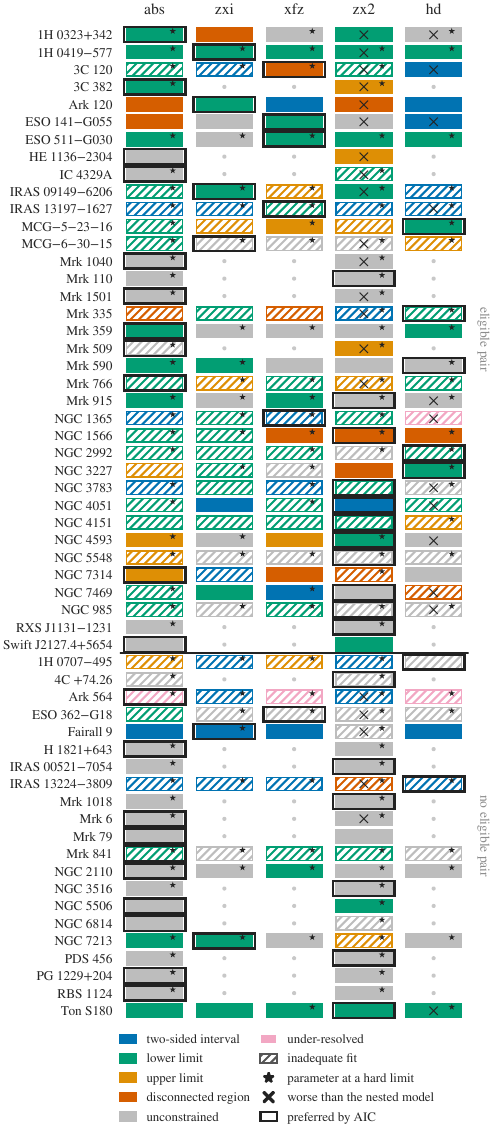}
\caption{Automated first pass (Appendix~\ref{app:audit}; not used for results):
type of spin region of each pipeline fit under its five variants, with
inadequate fits hatched, parameters at hard limits starred and the failures of
the exactly nested pair crossed.}
\label{fig:auditmatrix}
\end{figure*}

\paragraph{Adequacy.}  Of \Nfits\ pipeline fits, \Nacc\ have $p\ge0.01$, and
\Npatho\ place the inclination, iron abundance, ionisation, photon index or
reflection fraction at a hard limit (Figure~\ref{fig:auditmatrix}).

\paragraph{Sensitivity.}  The two-sided intervals that the pipeline found
(Fairall~9, Ark~120, NGC~4051) were not recovered when the pn data below 3~keV
were excluded, and those of Fairall~9 and Ark~120 occurred only at inclinations
of $77$--$87^\circ$.

\onecolumngrid

\bibliographystyle{aasjournal}
\bibliography{references_paper2}

\clearpage
\begin{table*}
\centering
\small
\caption{Published reflection spins from the compilations of
\citet[][R21, their Table~1]{Reynolds2021} and \citet[][SR26, their
Table~1]{SiskReynes2026}, and the reference fits of this analysis.  Mass in
$10^6\,M_\odot$ from R21 ($\sim$: approximate); ``not listed'': excluded from
SR26.  $^a$ SR26's Table~1 lists $>0.65$ for
MCG$-$6-30-15; their text adopts $\astar>0.93$.}
\label{tab:literature}
\begin{tabular}{llllll}
\toprule
Source & Mass & $\astar$ (R21) & $\astar$ (SR26) & This work & Status \\
\midrule
\csname @@input\endcsname tab_v9_literature.tex
\bottomrule
\end{tabular}
\end{table*}

\begin{table*}
\centering
\small
\caption{Reference model for every source.  Elig.: eligible pair; $p$:
nominal $\chisq$ probability; Fit: acceptable ($p\ge0.01$), exploratory or
inadequate; $\Delta\chisq_{\rm disc}$: increase of $\chisq$ when the disc
reflection is removed (detected if $>13.28$); Flags: N not at the minimum, P
implausible parameter, S disc reflection not detected or at a limit; $i$:
best-fit inclination (deg); Region: 90\% spin region; A: accepted.}
\label{tab:census}
\begin{tabular}{llllllllll}
\toprule
Source & Elig. & $\chisq/\nu$ & $p$ & Fit & $\Delta\chisq_{\rm disc}$ & Flags & $i$ & Region & A \\
\midrule
\csname @@input\endcsname tab_v9_census.tex
\bottomrule
\end{tabular}
\end{table*}

\begin{table*}
\centering
\footnotesize\setlength{\tabcolsep}{2.8pt}
\caption{Spin regions of the accepted sources (above the line) and, below the
line, of ESO~362$-$G18 and PDS~456 (refitted from their two-dimensional grid
minima, Section~\ref{sec:model}; their nested fits are from the earlier reference
fits), of Ark~120 (inclination at the limit)
and of Fairall~9 and Ton~S180 (no eligible pair),
under the reference model and the four nested changes.  Inclinations (deg) of
the reference fits in brackets.  $^\dagger$: fit not acceptable; $^\ast$:
flagged (N, P or S).  $\cup$ joins the pieces of a disconnected region; a bound
written $>x$ reaches the edge of the grid ($\astar\ge0.978$); $\approx$: a
piece narrower than 0.005; --: no 90\% region, because the profile never
returned to the best fit.}
\label{tab:arms}
\begin{tabular}{llllll}
\toprule
Source & reference & broken emissivity & $R_{\rm in}$ free & $\pm$Fe XXV/XXVI & x-cal slope \\
\midrule
\csname @@input\endcsname tab_v9_arms_a.tex
\bottomrule
\end{tabular}
\end{table*}

\begin{table*}
\centering
\footnotesize\setlength{\tabcolsep}{4pt}
\caption{As Table~\ref{tab:arms}, for the two sensitivity tests and the two
alternative models (inclinations in brackets).}
\label{tab:arms2}
\begin{tabular}{lllll}
\toprule
Source & $i\le60^\circ$ & pn $\ge3$~keV & high-density disc & lamppost \\
\midrule
\csname @@input\endcsname tab_v9_arms_b.tex
\bottomrule
\end{tabular}
\end{table*}

\begin{table*}
\centering
\footnotesize\setlength{\tabcolsep}{4pt}
\caption{Spin estimates from methods other than X-ray reflection for sources in
our sample, and the reference fits of this analysis.  Continuum: thin-disc
continuum fitting of the ultraviolet spectrum \citep{Capellupo2017}.
Microlensing: energy shifts of the microlensed Fe line \citep{Chartas2017}; the
value is suggested without a formal fit.  Outflow: jet-power method of
\citet{Daly2019}, who quote $\pm0.15$ for the sources of their Table~1 (shown
here) and no per-source uncertainty for the others; values of 1.00 are the
method's ceiling.  $i$: our best-fit inclination (deg).}
\label{tab:othermethods}
\begin{tabular}{lllllll}
\toprule
Source & Method & $\astar$ (other) & Reference & This work & $i$ & Status \\
\midrule
\csname @@input\endcsname tab_v9_othermethods.tex
\bottomrule
\end{tabular}
\end{table*}

\end{document}